\documentclass[twocolumn,10pt,a4paper]{article}

\usepackage{palatino}
\usepackage{amsmath}
\usepackage{amssymb}
\usepackage{graphicx}
\usepackage{graphics}

\begin{document}
\title{\sf \bf Sznajd Complex Networks}

\author{Luciano da Fontoura Costa \\
    Institute of Physics at S\~ao Carlos, \\
    Universidade de S\~ ao Paulo, \\
    S\~{a}o Carlos, SP, \\
    Caixa Postal 369, 13560-970, Brasil, \\
    luciano@if.sc.usp.br.  }

\maketitle

\begin{abstract}
The Sznajd cellular automata corresponds to one of the simplest and
yet most interesting models of complex systems.  While the traditional
two-dimensional Sznajd model tends to a consensus state (pro or cons),
the assignment of the contrary to the dominant opinion to some of its
cells during the system evolution is known to provide stabilizing
feedback implying the overall system state to oscillate around null
magnetization.  The current article presents a novel type of
geographic complex network model whose connections follow an
associated feedbacked Sznajd model, i.e. the Sznajd dynamics is run
over the network edges. Only connections not exceeding a maximum
Euclidean distance $D$ are considered, and any two nodes within such a
distance are randomly selected and, in case they are connected, all
network nodes which are no further than $D$ are connected to them.  In
case they are not connected, all nodes within that distance are
disconnected from them.  Pairs of nodes are then randomly selected and
assigned to the contrary of the dominant connectivity.  The topology
of the complex networks obtained by such a simple growth scheme, which
are typically characterized by patches of connected communities, is
analyzed both at global and individual levels in terms of a set of
hierarchical measurements introduced recently.  A series of
interesting properties are identified and discussed comparatively to
random and scale-free models with the same number of nodes and similar
connectivity.
\end{abstract}

\section{Introduction}

Introduced by Sznajd-Weron and her father Sznajd
in 2000~\cite{Sznajd}, the Sznajd cellular automata
represent one of the simplest and most interesting models of complex
systems.  Typically considered as a model of opinion formation, the
Sznajd model is known to undergo phase transitions in specific
circumstances~\cite{Sousa}.  Although recent, such a model has motivated
a whole series of investigations --- the interested reader should
refer to~\cite{forsta,AIP} for updated surveys.  While the traditional
two-dimensional Sznajd model is known always to reach consensus (all
pro or all cons) after its evolution, the assignment of the contrary
of the dominant opinion to some nodes during the evolutionary dynamics
is known~\cite{contra} to provide a stabilizing effect in the sense that
the system may now oscillate around null magnetization (i.e. balance of
pros and cons).  The two-dimensional spatial distribution of such
states involves a series of interconnected patches of different sizes
and shapes, which suggests the use of the Sznajd model for pattern
formation studies.

Also introduced recently, complex networks have motivated their own
focus of attention from the complex systems
community~\cite{AB:surv,Newman:surv}.  Representing a promising
interface between two well-established areas, namely graph theory and
statistical mechanics, this new area provides interesting
possibilities for representation, characterization and simulation of
systems with complex connectivity as well as the respective
dynamics~\cite{Newman:surv}.  Initiated by the pioneering studies by Flory
~\cite{Flory},
Rapoport~\cite{Rapoport} and Erd\"os and R\'enyi~\cite{Erdos_Renyi},
the area of complex networks has been catalysed by the more recent
developments by Watts and Strogatz~\cite{Watts_Strogatz} and
Barab\'asi and collaborators~\cite{Barabasi}. A great part of the
current interest in this area stems from scaling laws such as the
power-law characteristic of the Barab\'asi-Albert model, whose main
importance lies in the enhanced probability of obtaining hubs.  By
concentrating the network connectivity, such hub nodes have been found
to be of paramount importance for the network topology, dynamics, and
resilience to attack~\cite{AB:surv}.  Although complex networks are
typically characterized and analyzed in terms of the topological
measurements known as node degree and clustering coefficient, a set of
hierarchical measurements, including respective extensions of these
two concepts, was introduced
recently~\cite{PRL:Costa,Generalized,Hier_Char} which allows a
substantially more comprehensive characterization of the connectivity
of the analyzed networks.

The current article brings together the two above areas of Sznajd
models and complex networks.  Although Sznajd dynamics has already
been performed on such networks, we describe what is possibly the
first study involving Sznajd dynamics over the network connectivity.
More specifically, we suggest a novel geographic complex network model
whose connections are defined as a consequence of the evolution of a
respectively associated Sznajd model with contrary feedback.  The node
positions are distributed according to the Poisson density (i.e. the
nodes are positioned inside the $L \times L$ square with uniform
probability), and only pairs of nodes lying within Euclidean distance
$D$ one another are allowed to connect (see also~\cite{voronoi,topo})
during the subsequent processing stage.  While a model obeying such
rules would exhibit regular features, in the sense that the node
degrees of all nodes would be close to their overall average (small
node degree dispersion), the suggested model involves a dynamical
evolution of the connections by taking into account a respectively
associated feedbacked Sznajd model which implies enhanced node degree
dispersion as a consequence of border effects of the obtained patched
communities.  The initial network is obtained by assigning, with
uniform probability $p$, connections to each pair of nodes which are
no further apart than $D$.  The growth procedure involves selecting,
one at a time, a pair of nodes lying no further than $D$.  In case
these two nodes $i$ and $j$ are already connected, all the network
nodes which are up to the maximum Euclidean distance of $D$ from $i$
are connected to that node, and all nodes up to $D$ from $j$ are
connected to that other node.  In case the two nodes $i$ and $j$ are
disconnected, all network nodes which are no further than $D$,
respectively to each node $i$ and $j$, are disconnected from them.  In
this way, the `pros' and `cons' states in the traditional Sznajd model
are respectively associated to \emph{connection} and
\emph{disconnection}.

The complex networks obtained by the above described procedure
typically exhibit several interconnected patches of highly connected
nodes (communities), which is not unlike the distribution of
communication and energy in a town or country.  Such a distribution of
connections and communities provides an interesting prototype for
investigating the communication between nodes in the sense suggested
in~\cite{voronoi}, i.e. regarding the communication between nodes at
different positions.  Of particular interest is the
\emph{accessibility} between nodes belonging to distinct communities.
While limited information about such topological aspects can be
supplied by traditional measurements such as the node degree and
clustering coefficient~\cite{AB:surv,Newman:surv}, in this work we
apply the hierarchical measurements recently introduced
in~\cite{PRL:Costa,Generalized,Hier_Char}.  Such measurements, which
include hierarchical extensions of the node degree and clustering
coefficient, consider not only the immediate neighborhood of each
network nodes, but all the hierarchical levels defined by taking into
account each node as a reference.  As shown recently~\cite{Hier_Char},
such hierarchical measurements provide a comprehensive
characterization of the topology of traditional network models such as
random, scale-free (Barab\'asi-Albert --BA~\cite{Barabasi,AB:surv})
and geographic-regular (i.e. a mesh).  The application of such
hierarchical measurements to the Sznajd models introduced in this
article provides a series of interesting results, including the
existence of peaks of connectivity along the hierarchies, which are
illustrated and discussed, as well as similarities and dissimilarities
with the random and scale-free network models.

This work starts by describing the hierarchical measurements of
complex networks and follows by presenting the Sznajd geographical
complex network and its characterization in terms of hierarchical
topological measurements.

\section{Hierarchical Characterization of the Topology of Complex Networks} \label{sec:hier}

A non-oriented, non-weighted, complex network (or graph) with $N$
nodes and $E$ edges can be completely represented in terms of its
adjacency matrix $K$, such that $K(i,j)=1$ indicates the presence of a
connection between nodes $i$ and $j$, while absence of connections are
marked by null respective values in that matrix.  A complex network is
said to be \emph{geographical} in case each of its nodes has a
well-specified position in a metric space such as $R^2$.  An example
of geographical complex network is the Voronoi models described
in~\cite{voronoi}, as well as real networks heavily influenced by
adjacency constraints, such as transportation systems in the real
world.

Complex networks have often been characterized in terms of the average
node degree and clustering coefficient~\cite{AB:surv,Newman:surv}.
While the degree of a specific network node corresponds to the number
of edges attached to that node (observe that the degree of node $i$
can be immediately obtained by summing the entries along the column
$i$, or row, in the respective adjacency matrix), the clustering
coefficient can be defined in terms of the following equation

\begin{equation}
cc(i)= \frac{e(i)}{e_T(i)} = 2 \frac{e(i)}{n_1(i)(n_1(i)-1)} \label{eq:cc}
\end{equation}
where $e(i)$ is the number of edges among the immediate neighbors of
$i$ (connections with that node are not considered), $e_T(i)$ is the
maximum possible number of connections between those $n_1(i)$
neighbors.  Observe that $0 \leq cc(i) \leq 1$, with values of zero
being achieved for complete absence of connections and one for
complete connectivity among the $n_1(i)$ nodes.

Although the averages of node degree and clustering coefficient taken
over the whole network of interest provide important quantification of
the network connectivity, they are highly degenerated in the sense
that an infinite number of complex networks may present the same
values for those measurements.  The recently introduced concept of
hierarchical measurements~\cite{PRL:Costa,Generalized,Hier_Char}
provide valuable complementary information about the topological
properties of a network with respect to its several \emph{hierarchical
levels}.  Given a reference node $i$, hierarchies along the remainder
of the network can be established by considering the nodes which are
at successive exact distances, henceforth represented as $d$, from the
reference node $i$.  The edge distance $d$ between two nodes is
henceforth understood to correspond to the number of edges along the
shortest path between those two nodes~\footnote{It should be observed
that there are two distances considered in the current article: the
distance $d$ between nodes along network paths, measured in terms of
edges; and the Euclidean distance $D$ measured between the positions
of nodes in the Euclidean square $L \times L$}.  Interestingly, the
concepts of node degree and clustering coefficient can be immediately
extended to consider each hierarchical level with respect to the
reference node, and the respective averages over the network used for
a more comprehensive characterization of the connectivity of the
studied networks~\cite{Generalized,Hier_Char}.

Let us define the \emph{neighborhood} of a node $i$ at edge distance
$d$ as the set of nodes $R_d(i)$ which are exactly at distance $d$
from $i$, and let $\gamma_d(i)$ be the subnetwork defined by those
nodes plus the interconnections inherited from the original complete
network.  Such a subnetwork (and sometimes the set of nodes $R_d(i)$)
have been called the \emph{ring} of radius $d$ centered at the
reference node $i$~\cite{Generalized,Hier_Char}.  The
\emph{hierarchical degree} of node $i$ at edge distance $d$ can be
defined as the number of edges between the subnetworks $\gamma_d(i)$
and $\gamma_{d+1}(i)$.  As such, the hierarchical node degree at
distance $d$ ultimately considers the whole network of $i$ up to
distance $d$ as being a single enlarged node, whose degree is
naturally expressed as above.  The \emph{hierarchical clustering
coefficient} can be defined in analogous manner in terms of the
following expression

\begin{equation}
cc_d(i)= 2 \frac{e_d(i)}{n_d(i)(n_d(i)-1)}.
\end{equation}
where $e_d(i)$ corresponds to the number of edges among the nodes in
the subnetwork $\gamma_d(i)$ and $n_d(i)$ to the number of nodes in
that subnetwork.  As with the traditional clustering coefficient, we
also have that $0 \leq cc_d(i) \leq 1$, with analogous
interpretations.

Other hierarchical measurements which can be used to provide
additional information about the connectivity of the studied network
include the following:

\vspace{0.5cm}

{\bf Hierarchical number of nodes ($n_d(i)$):} the number of nodes in
the ring $\gamma_d(i)$, which is equal to the cardinality of $R_d(i)$.
This measurement has been found to be correlated with the hierarchical
node degree, although usually lagged ahead of that measurement.
Observe that the hierarchical number of nodes tend to be larger (or at
the most equal) to the hierarchical node degree because of the
\emph{convergence} of the edges while extending from one ring to the
next.

{\bf Intra-ring node degree ($A_d(i)$):} correlated to the
hierarchical node degree, this measurement provides the average degree
among the nodes of $\gamma_d(i)$.  Observe that only the edges between
the nodes in that subnetwork are taken into account.

{\bf Inter-ring node degree ($E_d(i)$):} corresponds to the average
degree of the nodes of $\gamma_d(i)$ considering only the connections
extending directly to the next ring $\gamma_P{d+1}(i)$.  Because the
nodes with higher degree (e.g. hubs) are more likely to appear in the
first hierarchical levels (they are more likely to be connected to the
reference node), this curve tends to decrease more steadly with $d$ in
case the network presents many hubs (e.g. scale-free).

{\bf Hierarchical common degree ($C_d(i)$):} the average of the
traditional node degree taken for each successive ring $\gamma_d(i)$.
This measurement has been found~\cite{Hier_Char} to provide good
discrimination for different network models, presenting a definetly
accentuated decrease for scale-free models and an interval of
sustained value for random and regular models.

\vspace{0.5cm}

Together with the hierarchical node degree and hierarchical clustering
coefficient, a total of 6 hierarchical measurements are therefore
obtained which are considered in the current article for the
characterization of the Sznajd geographical complex network.  Observe
that, given the finite size of real networks, \emph{all} such
hierarchical measurements tend to zero after some value of $d$, being
typically characterized by one or more peaks at characteristic values
of $d$.  An analytical expression for the hierarchical node degree
signature for random networks --- as well as illustration of the above
measurements for random, scale-free and regular network models, have
been described in ~\cite{Hier_Char}.  Such results indicate that the
number of hierarchical levels in a network tends to decrease for large
average node degree.

Given the patched nature of the complex networks obtained by the
herein suggested Sznajd complex model, the above described
hierarchical measurements stand out as particularly suitable for the
characterization of the intricate connectivity, including
bottlenecks, of the patched networks resulting from the Sznajd
dynamics over the network edges.

\section{The Sznajd Geographical Complex Network Model}

We now describe how a geographical complex network can have its
connections specified by the dynamical evolution of a feedbacked
Sznajd model.  The geographical network is assumed to be spatially
constrained within the square box of side $L$, while the nodes are
uniformly distributed along such space.  For the sake of improved
visualization, a minimal Euclidean distance $D_{min}$ is observed
between any two network nodes, therefore avoiding spatial
superposition of nodes.  Once the nodes are distributed, a network ---
henceforth called the \emph{underlying} network -- is obtained which
includes all connections between any two nodes which are no further
apart than a maximum Euclidean distance value $D_{max}$.  The
underlying network contains all connections which are possible to be
created or eliminated during the subsequent Sznajd dynamics. This
network is henceforth represented in terms of its adjacency matrix
$U$, whose respective number of edges is henceforth expressed as
$N_U$.

\begin{figure*}
 \begin{center} 
   (a) \includegraphics[scale=0.5,angle=-90]{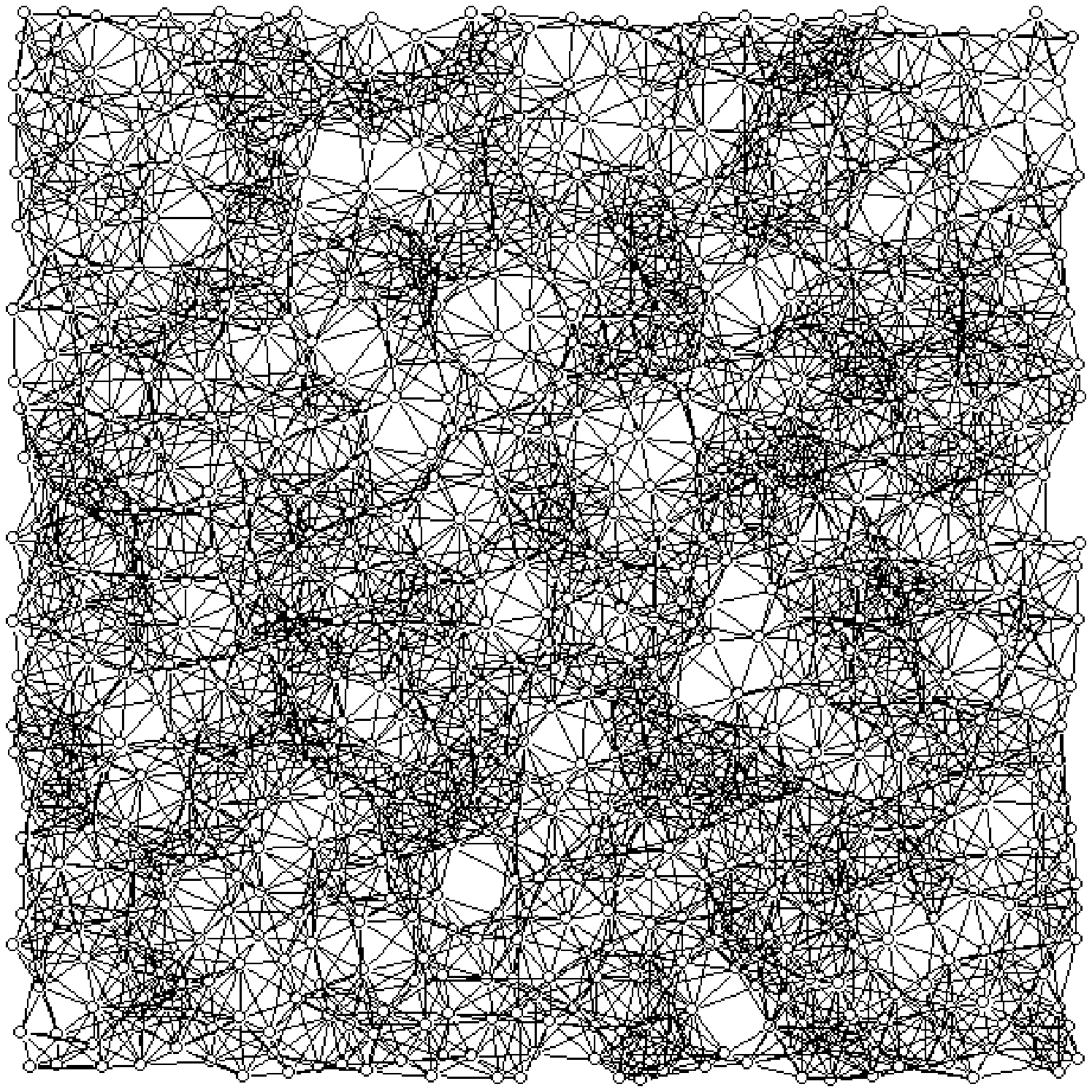} \\
   (b) \includegraphics[scale=0.5,angle=-90]{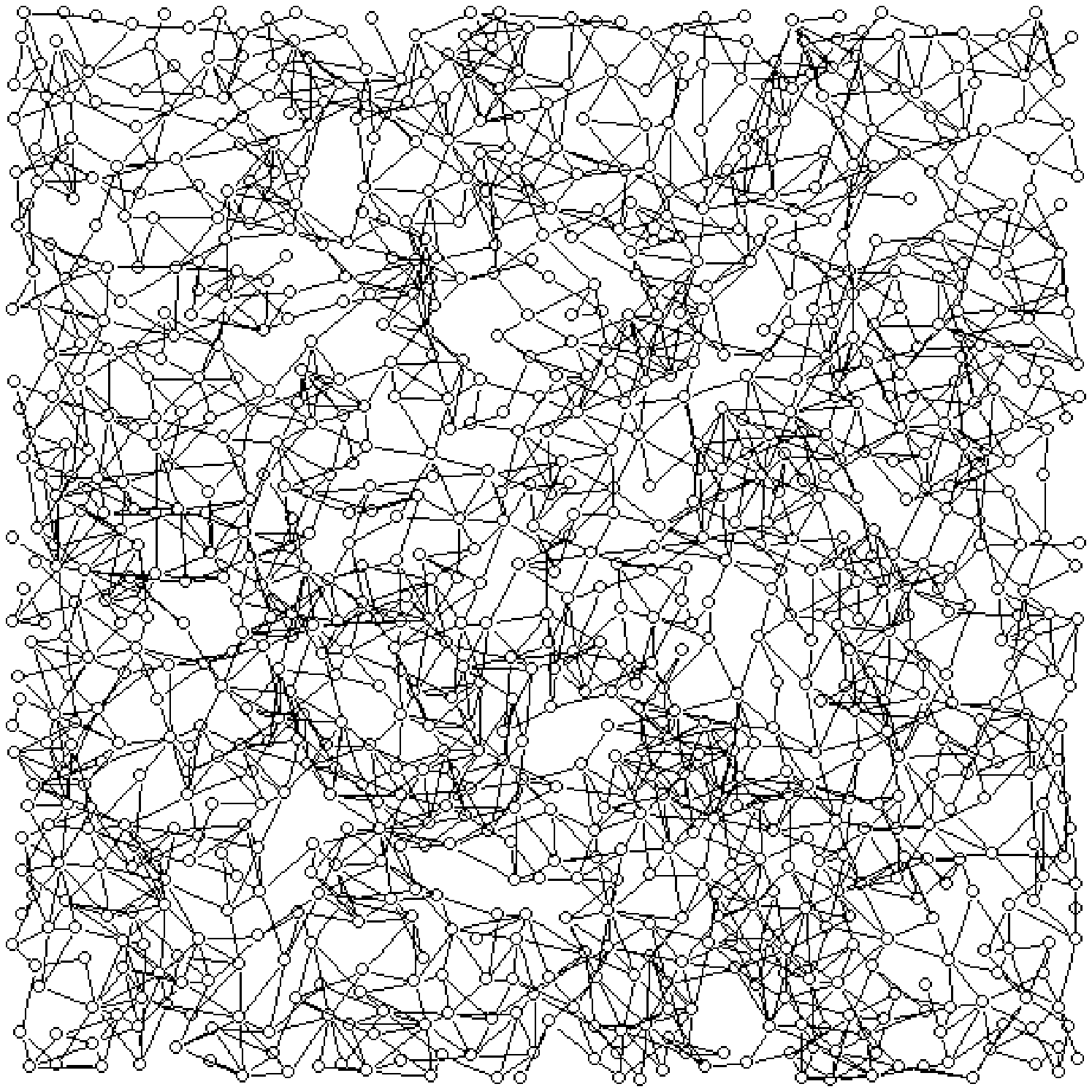}  \\
   (c) \includegraphics[scale=0.5,angle=-90]{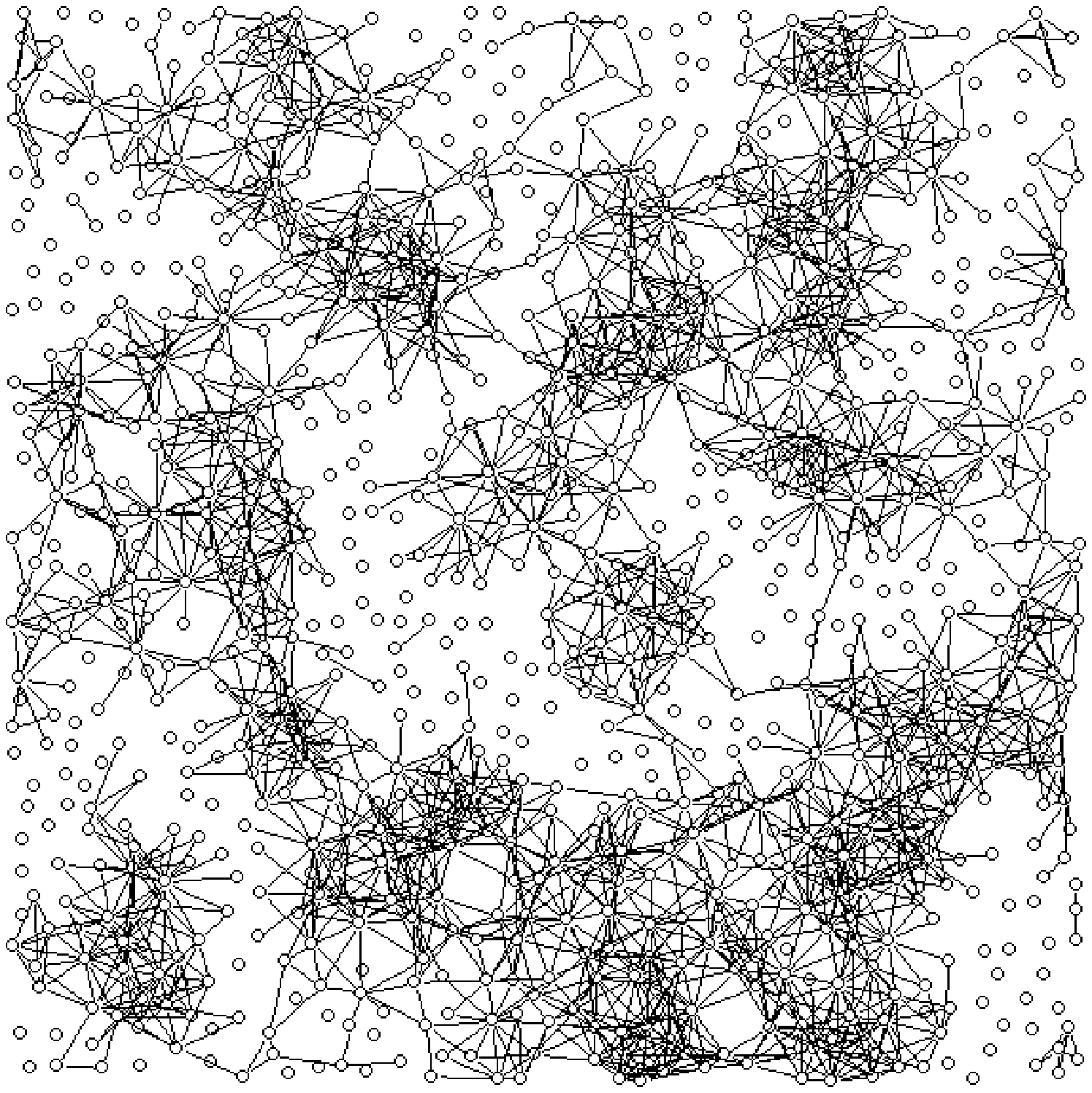}  \\
   \caption{Example of underlying network (a), a respectively derived
   initial configuration (b) and complex network obtained by using the
   feedback Sznajd dynamics (c).~\label{fig:ex}}
  \end{center}
\end{figure*}

Figure~\ref{fig:ex}(a) illustrates such a network obtained for
$N=1000$, $L=500$ and $D_{max}=35$.  Now, the \emph{initial
configuration} of the complex network (and Sznajd system) is obtained
by taking each of the edges in the potential network with uniform
probability $p$.  The initial complex network is henceforth
represented by its adjacency matrix $I$.  Figure~\ref{fig:ex}(b) shows
such a network obtained from the potential network in (a) by using
$p=0.5$.  The evolving network is represented by its adjacency matrix
$K$, with $K=I$ at the initial step.

Now, a feedbacked Sznajd dynamics is imposed on the above specified
initial configuration as follows.  An edge $(i,j)$ is sampled
uniformly from the adjacency matrix $U$ at a time.  Observe that such
edge may or may not exist in the currently evolving network
represented by $K$.  In case it exists (i.e. $K(i,j)=1$), all
immediate neighbors of $i$ and $j$ are identified and connected to
those two nodes.  In case the edge $(i,j)$ does not exist in the
current version of the evolving network (i.e. $K(i,j)=0$), all
immediate neighbors of $i$ and $j$ are disconnected from those two
nodes.  Observe that such a dynamics corresponds to the traditional
Sznajd model where the `pros' and `cons' opinions are now respectively
associated to \emph{connected} and \emph{disconnected}.

In case such a dynamics is allowed to take place for a long enough
period of time, the resulting network will be either a completely
connected or completely disconnected network, of little interest to
the complex network community.  More interesting connectivity can be
obtained by considering the Sznajd contrarian feedback scheme described
in~\cite{contra}.  Now, each time an edge is randomly sampled from $U$
and used to update the network as described above, another edge $e$
taken uniformly from $U$ is subsequently sampled with probability $q$
and the respective edge in $K$ receives the value contrary to the
current dominant `opinion' in the network.  For instance, in case the
current network represented by $K$ has more than $N_p/2$ edges, the
edge $e$ is disconnected in $K$.

After stabilization, which can be identified by observing the overall
mean magnetization to remain limited around zero during a
pre-specified interval of time, geographical complex networks are
obtained which are characterized by spatially delimitated patches of
highly connected communities which are themselves less intensely
interconnected one another.  Figure~\ref{fig:ex}(c) shows such a
network obtained from the initial condition in (b) by considering
$q=0.4$. Such networks are remindful of those resulting from the
distribution of energy and communication cables, or even social
relationships, in a town (where each community would correspond to a
building or institution) or a densely populated country (the
communities corresponding to the towns).  Because of such spatial
distribution of connections, the obtained networks are particularly
interesting to be used as prototypes for studies of communication and
accessibility between nodes in the sense discussed, for instance,
in~\cite{voronoi}.  Of particular interest is the number of
hierarchies taken for a specific node to broadcast along the network,
as well as the identification of the many bottlenecks.  Because of the
patched nature of the connections, it is expected that surges of
communication will take place every time a new highly connected
community is reached.

We show in the next section how the set of hierarchical measurements
of network topology recently introduced
in~\cite{PRL:Costa,Generalized,Hier_Char} can be used in order to
obtain a comprehensive characterization of intricate topology of
Sznajd geographical complex networks.

\section{Simulation Results}

The above obtained Sznajd complex network model was analyzed by using
the set of hierarchical measurements reviewed in
Section~\ref{sec:hier}.  Simulated random and BA network models with
the same number of nodes (1000) and similar connectivity ($\left< k
\right> \approx 5.5$) were also characterized by the same measurements
for the sake of comparison.  We divide the hierarchical
characterization of the networks into two subsections, one considering
the average $\pm$ standard deviation of the measurements considering
all network nodes, and another taking into account the hierarchical
measurements at individual node level.

\subsection{Global Features}

\begin{figure*}
 \begin{center} 
   \includegraphics[scale=1.5,angle=-90]{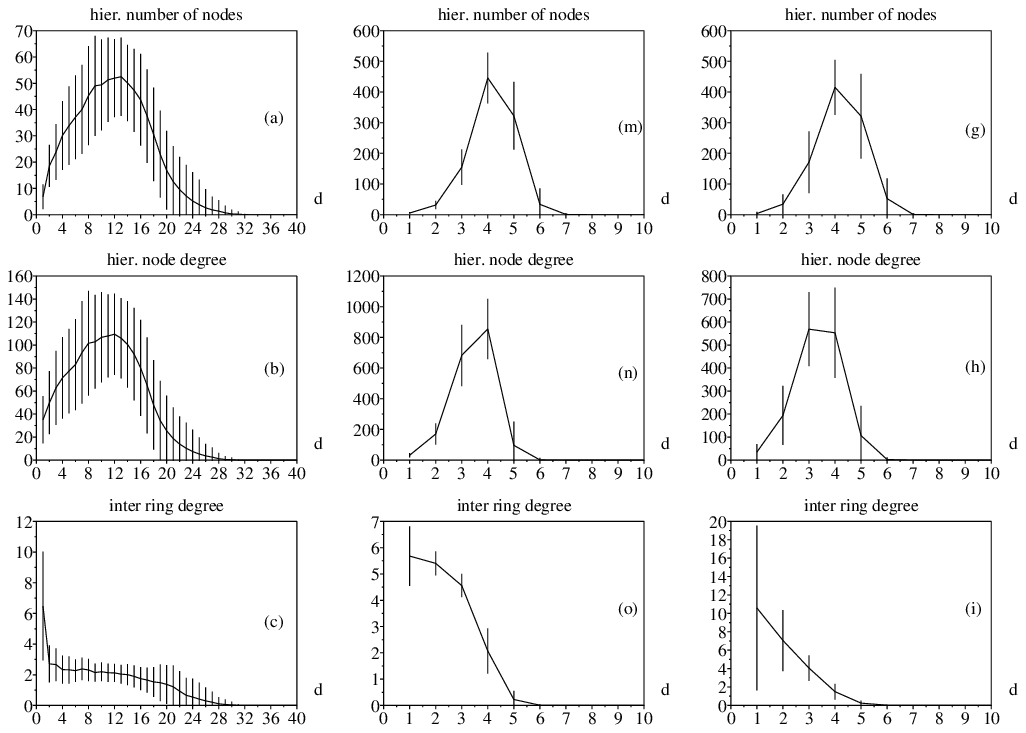} \\
   \includegraphics[scale=1.5,angle=-90]{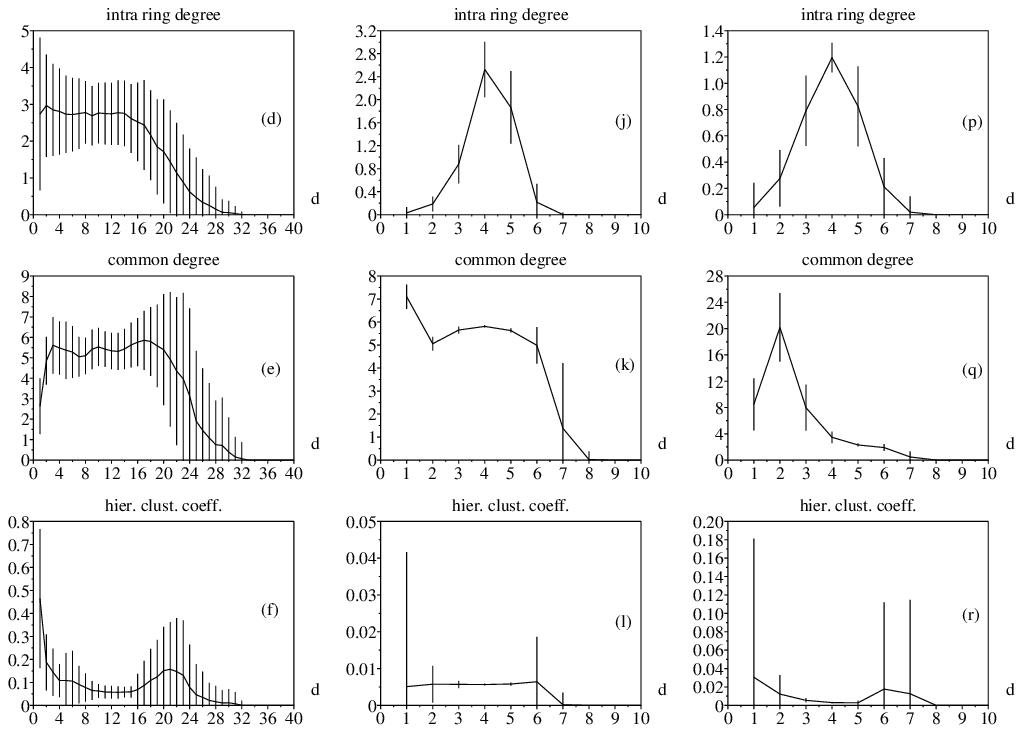} \\
   \caption{Hierarchical measurements obtained for the feedbacked
   Sznajd mode (first column) and random (second column) and BA (third
   column) simulated networks considering the same number of nodes and
   similar connectivity.~\label{fig:simuls}}
  \end{center}
\end{figure*}

Figure~\ref{fig:simuls} shows the average $\pm$ standard-deviations of
the set of hierarchical measurements obtained for the Sznajd
geographical network model (first column) as well as simulated random
and BA models considering the same number of nodes (1000) and similar
connectivity (5.5).  A marked difference is observed between the
Sznajd and the other models.  First, it includes many more
hierarchical levels, up to about 30 levels, while the number of
hierarchies in the other models is limited to about 10. As expected,
the shape of the curves obtained for the hierarchical number of nodes,
hierarchical node degree and hierarchical common degree involve three
distinct regions along the distance values $d$: (i) an increase
corresponding to the expansion of the neighborhood; (ii) a peak; and
(iii) a decrease implied by the limited size of the network.  When
observed comparatively one another, the shape of the curves obtained
for the hierarchical number of nodes and hierarchical node degree for
the three cases resulted similar, except for the respective heights
and widths.  The other measurements were characterized by similarities
and dissimilarities.  More specifically, the inter-ring degree and
hierarchical clustering coefficient of the Sznajd model (see
Figure~\ref{fig:simuls}(c) and (f), respectively) resembled --- in
shape, not absolute values --- the respective measurements obtained
for the BA model, as indicated in Figure~\ref{fig:simuls}(c) and (i).
At the same time, the hierarchical common degree of the curves
obtained for the Sznajd and random models resulted similar in shape
and absolute values.  The intra-ring degree of the former model was
not similar to any of the other two models.  Generally speaking, the
Sznajd network was characterized by sustained intra-ring degree (see
Figure~\ref{fig:simuls}(d)) along the initial distance values (roughly
between 1 and 15), which suggests that the rings obtained for that
model have similar interconnectivity.  A sustained interval was also
observed for the hierarchical common degree
(Figure~\ref{fig:simuls}(e)) and hierarchical clustering coefficient
(Figure~\ref{fig:simuls}(f)) obtained for the Sznajd network.  Such
sustained behavior are similar to those observed in~\cite{Hier_Char}
for regular networks, suggesting that the Sznajd model shares
topological features.  At the same time, the nearly constant values of
hierarchical node degree obtained for the Sznajd model
(Figure~\ref{fig:simuls}(d)) indicates that nodes with similar
traditional degrees are incorporated along about half of the
hierarchical levels, reflecting the absence of hubs in the Sznajd
model.  However the hierarchical clustering coefficient of the Sznajd
case was different in the sense that it contained a marked peak at
$d=22$ (see Figure~\ref{fig:simuls}(f)), which is more similar to
corresponding measurements obtained for the scale-free models.

\subsection{Individual Features}

\begin{figure*}
 \begin{center} 
   \includegraphics[scale=.7,angle=-90]{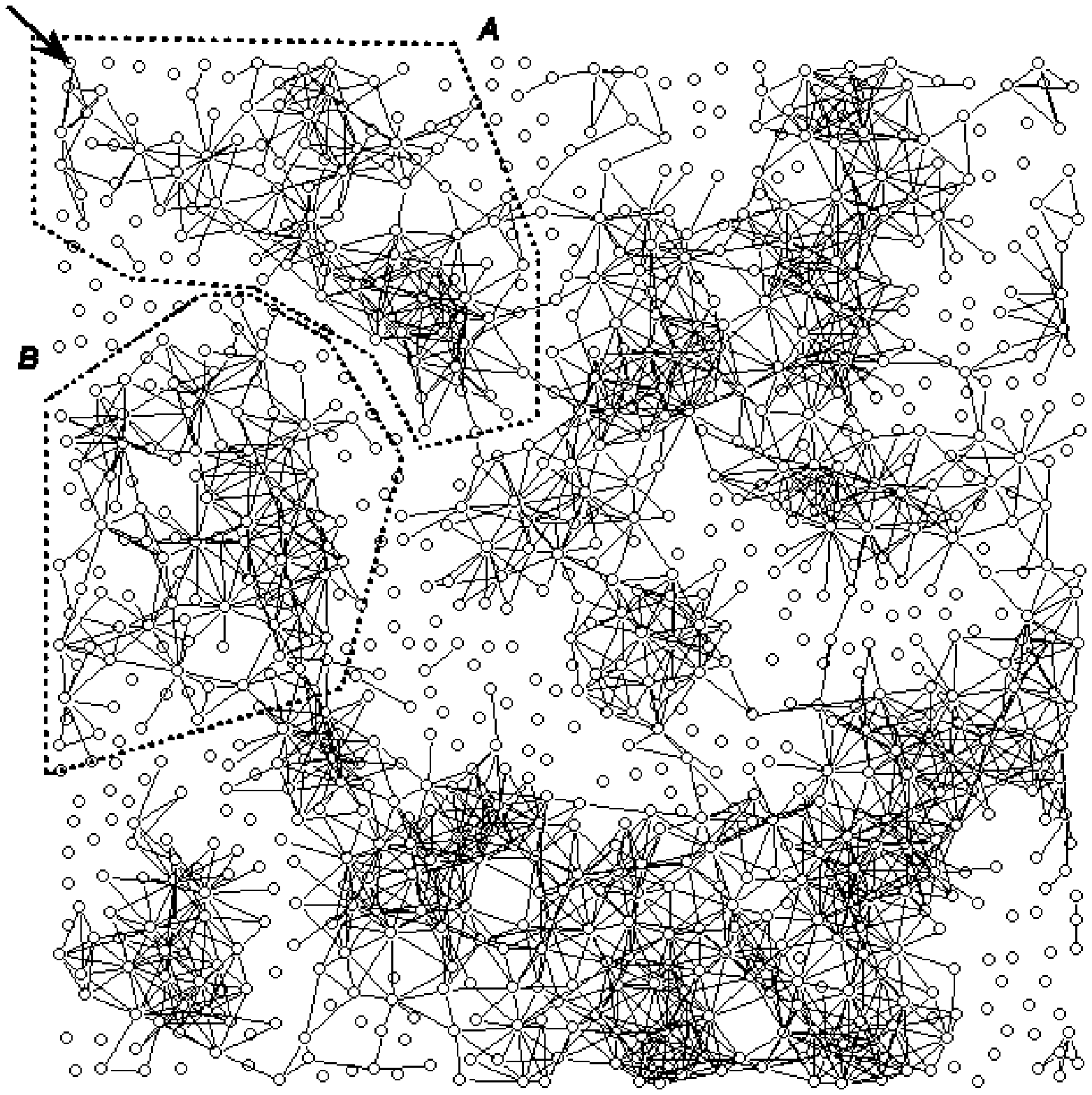} \\
   \caption{A modified version of the Sznajd network in
   Figure~\ref{fig:ex}(c), where the edges connecting the communities
   marked as {\bf A} and {\bf B} have been deleted for the sake of
   comparisons.  The network node considered for the individual
   hierarchical analysis is indicated by the arrow at the upper,
   rightmost corner of the network~\label{fig:communs}}.
  \end{center}
\end{figure*}

We now turn our attention to the analysis of the topological features
of the three considered models at an \emph{individual} level.  More
specifically, we concentrate our attention on the leftmost upper
network node, namely that indicated by the arrow in
Figure~\ref{fig:communs}.  This figure corresponds to the same network
in Figure~\ref{fig:ex}(c), except that the edges between the
communities {\bf A} and {\bf B} were deleted in the network shown in
Figure~\ref{fig:communs}, which was done in order to provide a
comparative context.

\begin{figure*}
 \begin{center} 
   \includegraphics[scale=1.5,angle=-90]{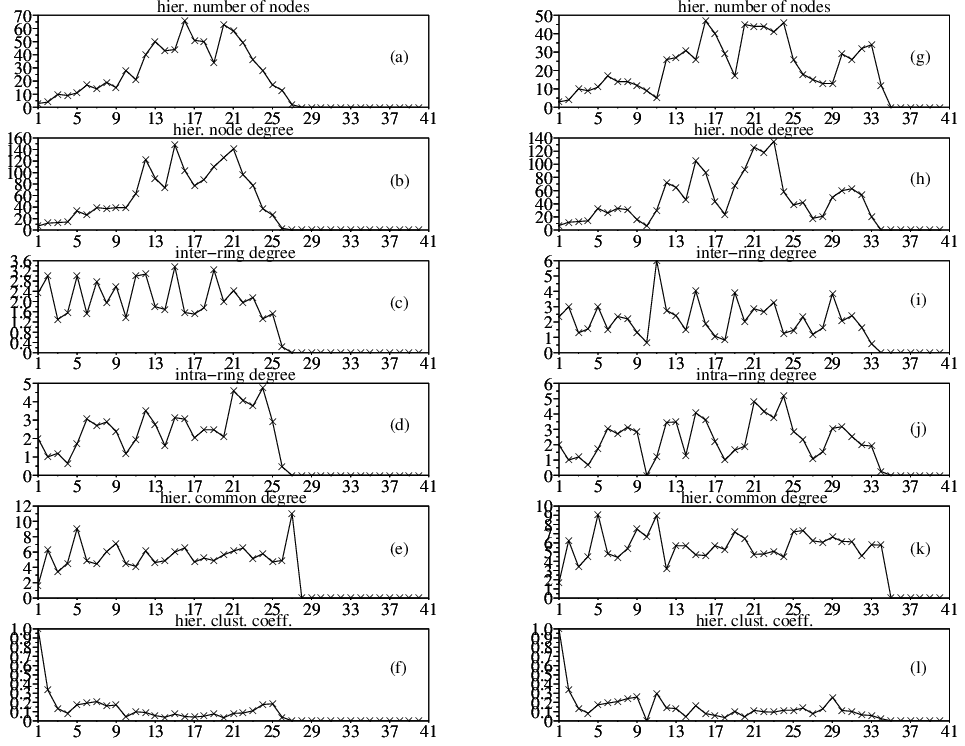} \\
   \caption{Hierarchical measurements obtained for the feedbacked
   Sznajd mode (first column) and random (second column) and BA (third
   column) simulated networks considering the same number of nodes and
   similar connectivity.~\label{fig:indiv}}
  \end{center}
\end{figure*}

Figure~\ref{fig:indiv} shows the hierarchical measurements obtained
for the Sznajd network in Figure~\ref{fig:ex}(c) (first column in
Figure~\ref{fig:indiv}) and the modified network in
Figure~\ref{fig:communs} (second column in Figure~\ref{fig:indiv}).
Several interesting features can be inferred from such results.
First, observe that all measurements obtained for the modified Sznajd
network extend along longer values of $d$, which is a consequence of
the fact that the original bypass from community {\bf A}, to which the
reference node (arrow) belongs, to community {\bf B} has been removed.
This small modification implied that the increasing hierarchies had to
go through the rest of the modified network before reaching the
community {\bf B}.  Actually, the peaks at the left and right hand
sides of the curve in Figure~\ref{fig:indiv}(g) --- i.e. the portions
of that curve extending from $d=1$ to 11 and from $d=29$ to 35,
respectively --- have been verified to correspond to the nodes in
communities {\bf A} an {\bf B}, respectively.

Another interesting feature is the similarity, except for a small
shift, between the curves obtained for the hierarchical number of
nodes and the hierarchical node degrees observed for both networks.
It has been verified that the valleys (i.e. minimum relative peaks)
along any of these curves correspond to the \emph{bottlenecks}
existing between the several spatially distributed network
communities.  Therefore, the identification of such minimum peaks
presents good potential for community finding in such networks, as
well as for any other network model.  The curves of the inter and
intra-ring degrees and hierarchical common degrees obtained for the
Sznajd network in Figure~\ref{fig:ex}(c) were characterized by
approximately sustained values, reflecting the degree regularity of
that model.  The counterpart curves obtained for the modified Sznajd
network (second column in Figure~\ref{fig:indiv}) were characterized
by being less regular, providing an illustration of the criticality of
small network changes upon the respective topological features.
Finally, the hierarchical clustering coefficients results similar for
both cases.

\section{Conclusions}

Unlike previous works~\cite{bergonz}, which ran Sznajd dynamics on complex
networks, we adopted the complementary approach and used the Sznajd
dynamics to produce complex networks.  The first point to be observed
is that the association between `pros' and `cons' with the
\emph{connected} and \emph{disconnected} states of edges,
respectively, can be used to define a broad Sznajd-related family of
complex networks.  In other words, the Sznaj dynamics is performed
over the network possible edges.

In the current work we explored the particular Sznajd dynamics where
randomly selected edges are assigned the opposite of the current
dominant connectivity among the evolving network.  The obtained
geographical networks were characterized by spatial patches of high
connectivity (i.e. communities) which were less intensely connected
one another.  Such models, which are remindful of distribution of
intercommunication, energy or even social contacts in real structures
such as towns and countries, provide an interesting prototype for
studying community and connectivity in complex networks.

Several interesting topological properties of the Sznajd networks were
identified by using a recently introduced set of hierarchical
measurements including hierarchical extensions of the traditional node
degree and clustering coefficient.  The analysis of the Sznajd model
was performed comparatively with random and scale-free (BA) models
simulated with the same number of nodes and similar connectivity and
by considering global measurements (i.e. average and standard
deviation) involving all network nodes as well as at the individual
node measurement level.  A series of interesting results was obtained,
including the identification of similarities and dissimilarities of
specific measurements between the Sznajd and random/BA models.
Generally, the Sznajd model was characterized by sustained values of
inter and intra-ring degrees, as well as hierarchical common degree
and hierarchical clustering coefficient.  The latter measurement also
resulted similar to that obtained for the BA case.  The Sznajd model
was globally characterized as exhibiting high node degree regularity,
a consequence of the spatial adjacencies underlying that model.  The
individual analysis of the hierarchical features was performed
considering a modified version of the Sznajd model, which lead to
major effects over the respective measurements.  Such examples clearly
illustrated the potential of the hierarchical measurements for
identification of communities and connectivity bottlenecks.

The above reported developments pave the way to a series of future
developments, including alternative growth network schemes and
applications.  For instance, it would be interesting to consider
Sznajd dynamics involving long-range connections~\cite{Schulze} as
well as higher spatial dimensions~\cite{bergonz}.  Other possibilities
involve the use of Sznajd geographical networks as models of
biological pattern formation, including neuronal modules/systems and
gene expression and cell signaling.  It would also be interesting to
develop a more systematic investigation of the potential of the
hierarchical measurements for identification of communities in complex
networks.

\vspace{0.5cm}
{\bf Acknowledgment:} Luciano da F. Costa is grateful to FAPESP
(proc. 99/12765-2), CNPq (proc. 308231/03-1) and the Human Frontier
Science Program (RGP39/2002) for financial support.

\end{document}